\documentclass[preprint,12pt]{elsarticle}
\usepackage{amssymb}
\usepackage{amsmath}
\usepackage{setspace}
\usepackage{graphicx}
\usepackage{lineno}

\begin{document}

\begin{frontmatter}

\title{Identical inferences about correlated evolution arise from ancestral state reconstruction and independent contrasts}
\author{Michael G. Elliot\corref{telephone}}
\ead{micke@sfu.ca}
\address{Department of Biological Sciences, Simon Fraser University,\\British Columbia, Canada, V5A 1S6}


\begin{abstract}
Inferences about the evolution of continuous traits based on reconstruction of ancestral states has often been considered more error-prone than analysis of independent contrasts. Here we show that both methods in fact yield identical estimators for the correlation coefficient and regression gradient of correlated traits, indicating that reconstructed ancestral states are a valid source of information about correlated evolution. We show that the independent contrast associated with a pair of sibling nodes on a phylogenetic tree can be expressed in terms of the maximum likelihood ancestral state function at those nodes and their common parent. This expression gives rise to novel formulae for independent contrasts for any model of evolution admitting of a local likelihood function. We thus derive new formulae for independent contrasts applicable to traits evolving under directional drift, and use simulated data to show that these directional contrasts provide better estimates of evolutionary model parameters than standard independent contrasts, when traits in fact evolve with a directional tendency.
\end{abstract}

 
\end{frontmatter}

\spacing{2}

\section*{Introduction}

Statistical methods for the detection of correlated evolution have been divided into two broad classes. Directional methods involve reconstruction of ancestral states followed by statistical inference based on the deviation in trait values along each branch of a phylogenetic tree, while nondirectional or cross-sectional methods involve comparisons of trait values across taxa rather than along branches \citep{HarveyPagel1991,Pagel1993}. Methods arising from the Brownian motion model, in which traits evolve over time by accumulating increments drawn from a symmetrical zero-centred distribution with fixed variance, include both directional approaches such as reconstruction of ancestral states under maximum likelihood or squared-change parsimony criteria \citep{Maddison1991,Pagel1993,McArdleRodrigo1994,Schluteretal1997} and nondirectional approaches such as independent contrasts \citep{Felsenstein1985,Garlandetal1992} and phylogenetic generalized least squares \citep{Grafen1989,MartinsHansen1997}.

It is well known that all methods based on the Brownian motion model are ultimately means of estimating the same model parameter, namely the variance of the Brownian process underlying trait evolution \cite{Pagel1993,Pagel1997,Rohlf2001,Freckleton2012}. The mean squared standardized independent contrast across the internal nodes of a phylogeny is an estimator of this parameter, while the mean squared deviation of reconstructed trait value across the branches of a phylogeny is an estimator of half this parameter \citep{Ackerly2009}. The close association of methods based on Brownian motion is further indicated by the facts that the phylogenetic mean trait value inferred under indendent contrasts is identical to the global maximum likelihood estimate of the root's trait value \citep{Garlandetal1999,Revelletal2008}, that independent contrasts and phylogenetic generalized least squares models yield identical regression estimators for the slope and gradient of two correlated traits \citep{Blombergetal2012}, and that regression coefficients of bivariate data estimated under directional and nondirectional approaches are highly correlated \citep{Pagel1997}.

The primary reason to select one class of method over another is thus not that they measure different things but that their estimators exhibit different statistical properties that may be more or less desirable \citep{Pagel1993}. In this sense, independent contrasts and phylogenetic generalized least squares models are generally favoured over ancestral state reconstruction. \citet{Pagel1993} argues that independent contrasts are best suited to the problem of identifying evolutionary correlation coefficients, since directional methods based on a tree with $n$ tips count evolutionary changes on $2(n-1)$ internal branches, meaning that ``half of the variation that a directional method calculates is redundant because it overlaps with variation already calculated'' yielding ``results that seem more stable than they actually are'', whereas independent contrasts, based on values calculated at $n-1$ internal nodes, ``make use of all the variance in the data, but in a way that does not count any of it twice''. \citet{Ackerly2009} concurs with this view and adds that deviations in trait value occurring on internal branches of a phylogeny are not independent, since trait deviations associated with each sibling pair of branches depend on the value of the ancestral state at the pair's common ancestor. Based on an analysis of phenotypic change in a bacteriophage colony with known evolutionary history, \citet{OakleyCunningham2000} advocate ``the use of independent contrasts in addition to or instead of the more error-prone ancestral estimation procedures'', error they ascribe to the existence of a directional bias in the polarity of trait change over time in their dataset. Directional tendencies in the evolutionary process have been shown to reduce the accuracy of ancestral state estimation in studies using fossil calibration to assess reconstruction quality \citep{FinarelliFlynn2006} and the quality of ancestral state reconstruction has been challenged in general \citep{Donoghueetal1989,WebsterPurvis2002,Slateretal2012}.

It is shown here that independent contrasts and maximum likelihood ancestral state reconstruction not only estimate the same underlying Brownian rate parameter for a univariate trait, but also -- in studies of correlated evolution -- yield numerically identical regression estimators for the gradient and correlation coefficient of bivariate traits. As a consequence, inferences about correlated evolution derived from maximum likelihood ancestral state estimation are as valid as, and indeed identical to, those derived from independent contrasts procedures. We show that the independent contrast associated with a pair of sibling nodes in a phylogenetic tree can be expressed in terms of the Gaussian local likelihood function of the node that is the direct common ancestor of the pair. It thus transpires that the numerical calculations carried out in generating independent contrasts are identical to those carried out in maximum likelihood ancestral state estimation in both univariate and multivariate situations. One consequence of this finding is that novel formulae for independent contrasts can be derived for any model of trait evolution for which a local likelihood function can be defined, including non-standard models that deviate from classical neutral assumptions. As a demonstration we derive new formulae for independent contrasts appropriate for a Brownian motion model of trait evolution with directional drift, which, in a bivariate context, are shown to yield more accurate estimates of correlation coefficient and slope than standard independent contrasts when the underlying evolutionary process does in fact exhibit a directional tendency. These findings are discussed in the context of claims that ancestral state esimation is in some sense more error-prone than independent contrasts. 

\section{Methods}

Our primary results depend on the standard Brownian motion likelihood function for a trait $\textbf{X}$ evolving over a rooted bifurcating phylogenetic tree such that the deviation in trait value along a branch of length $t$ is normally distributed with variance proportional to $t$. Our formulae refer to a general node $n$ whose child nodes are denoted $i$ and $j$ connected by branches of length $t_i$ and $t_j$ respectively, and whose parent node $p$ is connected by a branch of length $t_p$. Trait $\textbf{X}$ takes value $x_n$ at node $n$. The likelihood of an ancestral state assignment is given by: 
\begin{equation}
\textrm{L}(\textbf{X}; \textbf{T}) \propto \prod_{n} \phi(x_n-x_{p};0,\sqrt{t_n})
\end{equation}
where $\phi(x;\mu,\sigma)$ is the density of the Gaussian distribution $\textrm{N}(\mu,\sigma)$ evaluated at $x$. Each node $n$ is associated with a Gaussian global maximum likelihood function which describes the maximized likelihood of the tree conditional on the value of $x_n$, denoted $\textrm{N}(\hat{\mu},\hat{\sigma})$, and also with a Gaussian local maximum likelihood function describing the maximized likelihood of the subtree rooted at $n$ condition on the value of $x_n$, denoted $\textrm{N}(\tilde{\mu},\tilde{\sigma})$.

In our results we derive new formulae for independent contrasts accommodating traits evolving under Brownian motion with directional drift. In order to assess the performance of these directional independent contrasts in comparison with standard independent contrasts in identifying the slope and correlation coefficient of a pair of continuous characters evolving with a directional tendency, simulation studies were performed on one thousand random Yule trees, each with a number of trips drawn uniformly from 40 to 400. For each tree, evolution was simulated under a bivariate Brownian motion model with random reduced major axis regression slope (drawn uniformly from 0.2--2), random correlation coefficient (drawn uniformly from 0.2--1) and random drift parameters $M_{\textbf{X}}$ and $M_{\textbf{Y}}$ (drawn uniformly from 0--$2\sigma^2_{\textbf{X}}$ and 0--$2\sigma^2_{\textbf{Y}}$). The realized RMA regression slope and correlation coefficient were recorded, and then re-estimated, on the basis of tips data only, using standard independent contrast and directional independent contrasts as defined in Equation \ref{eq:directionalIC} below.

\section{Results} 

Supplement S1 shows that the standardized independent contrast between nodes $i$ and $j$, $\textrm{IC}_{(i,j)}$ can be expressed as the sum of the squared directional (\textit{sensu} \citet{Pagel1993}) standardized deviations in local maximum likelihood ancestral state as follows:
\begin{equation}
  \textrm{IC}_{(i,j)}^2 = \dfrac{(\tilde{\mu}_i-\tilde{\mu}_n)^2}{t_i + \tilde{\sigma}_i^2} + \dfrac{(\tilde{\mu}_j-\tilde{\mu}_n)^2}{t_j + \tilde{\sigma}_j^2}\label{eq:IC}
\end{equation}

An estimator for the variance of an evolving trait $\textbf{X}$ based on global maximum likelihood ancestral state reconstruction is given by:
\begin{equation}
  \textrm{var}[\textbf{X}] = 2 \textrm{E} \left[ \dfrac{ (\hat{\mu}_n - \hat{\mu}_p)^2 }{t_p} \right]
\end{equation}
while the covariance of traits $\textbf{X}$ and $\textbf{Y}$ is given by 
\begin{equation}
  \textrm{cov}[\textbf{X},\textbf{Y}] = 2 \textrm{E} \left[ \dfrac{ (\hat{\mu}_{\textbf{X}n} - \hat{\mu}_{\textbf{X}p})(\hat{\mu}_{\textbf{Y}n} - \hat{\mu}_{\textbf{Y}p}) }{t_p} \right]
\end{equation}
Supplement S2 demonstrates that these variance and covariance estimators are numerically identical to the variance and covariance of the set of independent contrasts generated from the same phylogeny and data.

An estimator for the reduced major axis (RMA) regression slope between $\textbf{X}$ and $\textbf{Y}$ based on maximum likelihood ancestral state reconstruction is given by
\begin{equation}
 |\beta| = \sqrt{ \dfrac{\textrm{E} \left[ (\hat{\mu}_{\textbf{Y}n} - \hat{\mu}_{\textbf{Y}p})^2 /t_p \right]}{\textrm{E} \left[(\hat{\mu}_{\textbf{X}n} - \hat{\mu}_{\textbf{X}p})^2/t_p \right]}}
\end{equation}
Supplement S3 demonstrates that this regression gradient estimator is numerically identical to the RMA regression estimator based on independent contrasts, and that this identity also holds for ordinary least squares regression.

Equation \eqref{eq:IC} can be used to generate formulae for independent contrasts appropriate for any model of trait evolution for which a local likelihood function can be defined. Supplement S4 derives formulae for a model of trait evolution with a directional tendency:
\begin{equation}
 \textrm{directional IC}_{(i,j)} = \dfrac{(\tilde{\mu}_i-\tilde{\mu}_j)- M (t_i - t_j)}{\sqrt{t_i+\tilde{\sigma}_i^2+t_j+\tilde{\sigma}_j^2}}\label{eq:directionalIC}
\end{equation}
where $M$ is the mean directional drift per unit time (with $M=0$ under standard independent contrasts) and where $\tilde{\mu}_i$ and $\tilde{\sigma}^2_i$ are estimated recursively from the tips to the root of the phylogeny according to
\begin{equation}
 \tilde{\mu}_n = \dfrac{(\tilde{\mu}_i-t_iM ) (\tilde{\sigma}_j^2 + t_j) + (\tilde{\mu}_j -t_jM )(\tilde{\sigma}_i^2 + t_i)}
 { (\tilde{\sigma}_i^2 + t_i) +  (\tilde{\sigma}_j^2 + t_j)}\label{eq:directionalIC2}
\end{equation}
\begin{equation}
 \tilde{\sigma}_n=\sqrt{\dfrac{(\tilde{\sigma}_i^2 + t_i)(\tilde{\sigma}_j^2 + t_j)}{(\tilde{\sigma}_i^2 + t_i)+(\tilde{\sigma}_j^2 + t_j)}}\label{eq:directionalIC3}
\end{equation} 
Supplement S4 also includes formulae for calculating the maximum likelihood ancestral state reconstruction under this model. The parameter $M$ is typically not known \textit{a priori} but is easily estimated by a linear search maximizing the likelihood of an ancestral state assignment (or, identically, minimizing the sum of squared contrasts). Results of simulation studies comparing performance of directional independent contrasts to standard independent contrasts in estimating correlation coefficient, RMA gradient and $M$ are illustrated in Figures 1 and 2. While the directional model has generally been regarded as underidentified \citep{Grafen1989} we find maximum likelihood estimates of $M$ to be identified for phylogenies that are not perfectly balanced in both topology and tips data. Software for maximum likelihood estimation of $M$ along with standardized directional independent contrasts has been made available at http://www.sfu.ca/\textasciitilde micke/dirpic.html.

\section{Discussion}

Maximum likelihood ancestral state reconstruction has often been regarded as a poor second cousin to nondirectional analysis of correlated evolution using independent contrasts of phylogenetic generalized least squares \citep{Pagel1997,OakleyCunningham2000}. Ancestral reconstruction has been regarded as more error-prone \citep{OakleyCunningham2000}, requiring fossil calibration to improve accuracy in the reconstruction of directional deviations in trait value \citep{Donoghueetal1989,WebsterPurvis2002,Slateretal2012}. Independent contrasts, being nondirectional, have been considered more robust with respect to such sources of bias and error. In this paper we have shown that, to the contrary, regression estimators based on maximum likelihood ancestral state reconstruction are numerically identical to estimators based on independent contrasts. Previous authors have considered the calculation of ancestral states using independent contrasts to yield identical results as direct methods such as those of \citet{Schluteretal1997} but ``without the use of maximum likelihood''. We have shown that the numerical calculations involved in calculating independent contrasts are in fact identical to those involved in fitting the maximum likelihood model and that the standardized independent contrast associated with a pair of nodes of a phylogeny can be expressed directly in terms of the Gaussian likelihood function at those nodes and their common parent. It has previously been proposed that maximum likelihood estimates yield too narrow confidence intervals, since there are twice as many branches in a phylogeny than there are internal nodes \citep{Pagel1993,Pagel1997}. We agree with \citep{Ackerly2009} that this apparent overconfidence is wholly remedied by reducing the number of degrees of freedom in the calculation of confidence bounds by a factor of two when using ancestral state reconstruction, though this manipulation is not necessary to guarantee the identity of point estimates made by regression estimators under 
ancestral state reconstruction and independent contrasts. In the light of findings that regression estimators based on independent contrasts are also identical to those based on phylogenetic generalized least squares \citep{Blombergetal2012}, we conclude that all comparative methods based on the Brownian motion model of evolution yield identical inferences about the parameters of correlated evolution and are conceptually indistinguishable. Our response to claims that ancestral state reconstruction is error-prone is to point out that the ancestral states themselves are merely nuisance parameters of the model formulation. Cross-sectional methods simply embed this error into the values of independent contrasts themselves. In estimating summary statistics of these nuisance parameters, such as evolutionary rate or correlation coefficient, independent contrasts offers no advantage over ancestral state-based methods.

One useful implication of this model is that ancestral states under non-standard models of trait evolution contain useful information about correlation structure. Given the evidence that directional tendencies give rise to biased estimates of evolutionary model parameters, it may be useful to incorporate such tendencies directly into the model likelihood function. For those wedded to the idiom of independent contrasts, we have shown that Equation \ref{eq:IC} can be used to generate novel formulae for independent contrasts when an appropriate likelihood function can be formulated. Specifically, we here present formulae for independent contrasts under directional drift (Equations \ref{eq:directionalIC}--\ref{eq:directionalIC3}) and show that these ``directional independent contrasts'' markedly improve estimation of correlation coefficient and slope (Figures 1 and 2). 

More generally, ancestral state reconstructions of traits evolving under wholly non-Brownian statistical models \citep{Landisetal2013,ElliotMooers2013}, which entirely invalidate the assumptions of independent contrasts and phylogenetic generalized least squares as currently formulated, still contain useful information about correlated evolution. Maximum likelihood fitting of ancestral states is a useful general strategy for complex models of trait evolution, it is appropriate to use these reconstructed states to make inferences about historical patterns and processes of correlated evolutionary change when cross-sectional methods such as independent contrasts are not available.

\section*{Acknowledgements}

The author thanks NSERC and the Human Evolutionary Studies Program at Simon Fraser University for financial support, and members of the FAB* and HESP groups for their comments.

\bibliographystyle{model1-num-names}
\bibliography{refs}{}\underline{}

\pagebreak

\begin{figure}[ht!]
   \includegraphics[width=1\textwidth]{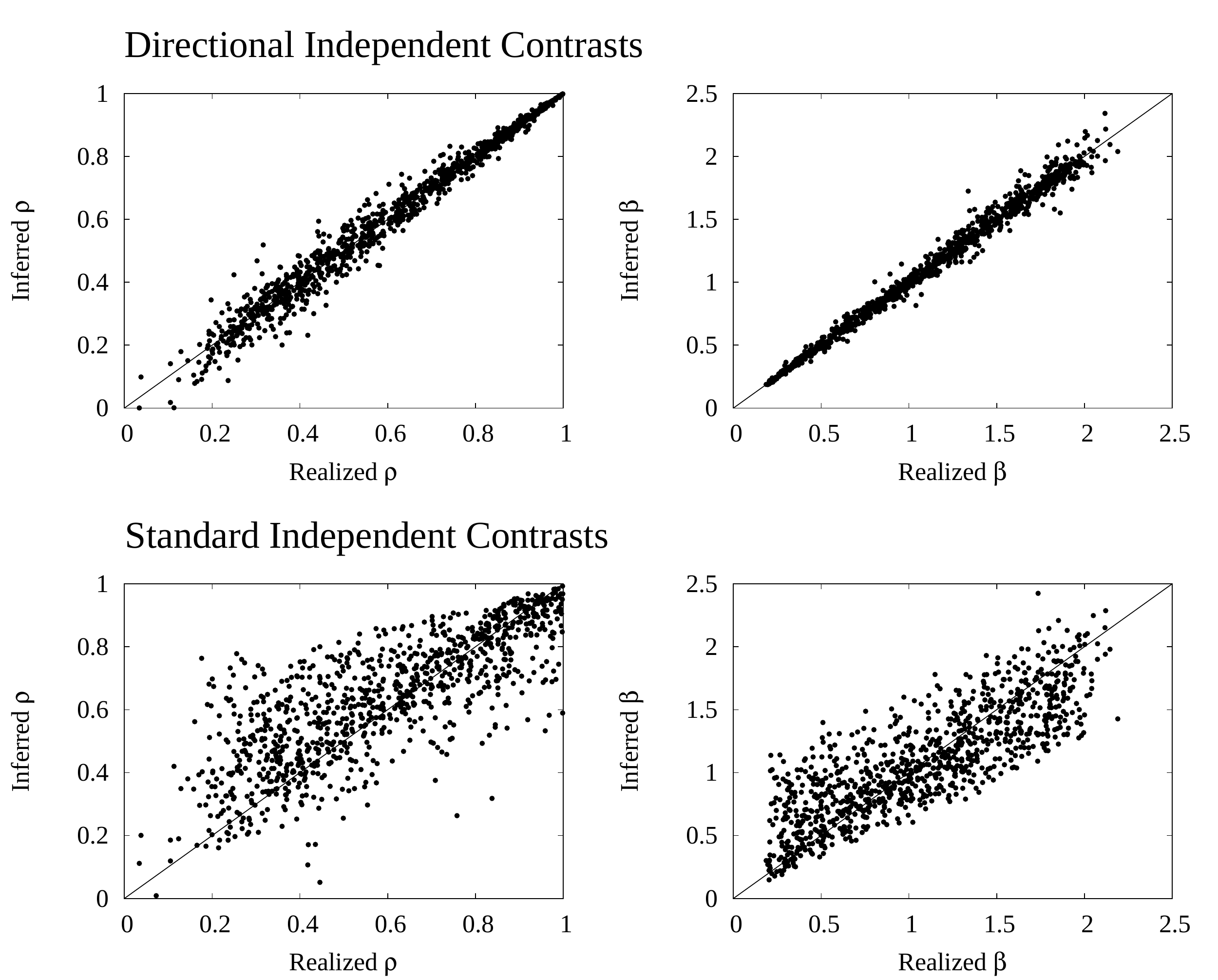}
   \caption{Accuracy in estimation of correlation coefficient ($\rho$, left column) and regression slope ($\beta$, right column) of two traits evolving under bivariate Brownian motion with directional drift, using directional independent contrasts (top row) and standard independent contrasts (bottom row). The reference line for perfect estimates is included on each panel.}
\end{figure}

\pagebreak

\begin{figure}[ht!]
   \centering
   \includegraphics[width=0.5\textwidth]{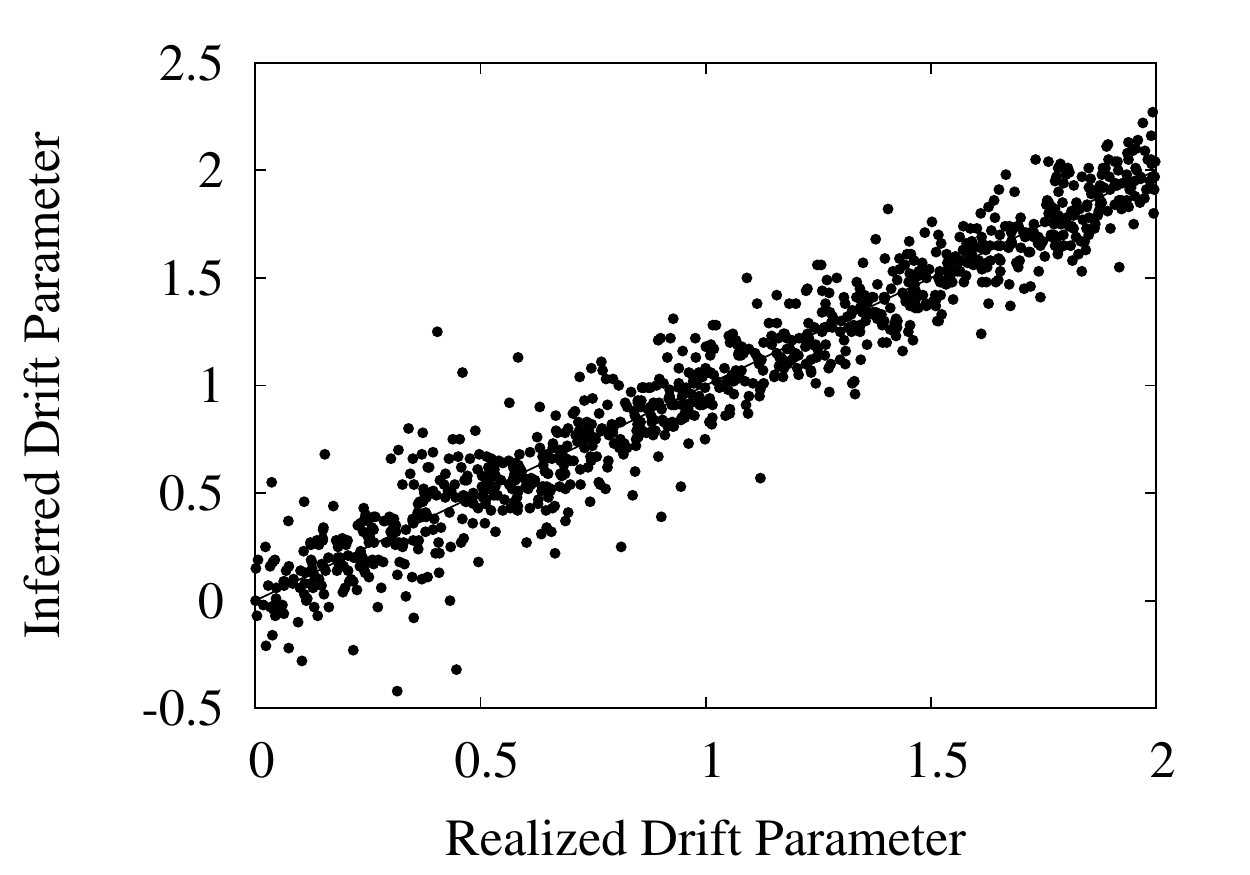}
   \caption{Accuracy in estimation of the directional drift parameter under directional independent contrasts.}
\end{figure}

\pagebreak

{\Large Supplementary Information}
\setcounter{section}{0}

\section{Introduction}

\subsection{Terminology}

We consider a continuous biological trait $\textrm{X}$ evolving over a rooted bifurcating phylogenetic tree. At each node $n$, the biological trait has value $x_n$. Each node $n$ has two descendant nodes, $i$ and $j$, unless $n$ is a tip. Each node $n$ has a parent node $p$, unless $n$ is the root. The branch connecting node $n$ to its parent has length $t_n$. 

\subsection{Independent contrasts}

Independent contrasts are calculated using an algorithm that traverses the phylogeny from tips to root, iteratively calculating transformed trait values $\textrm{X}^\prime$ and transformed branch lengths $\textrm{T}^\prime$ along the way. Following \citet{Felsenstein1985} the independent contrast associated with node $n$, $\textrm{IC}_n$, is defined as:
\begin{equation}
\label{eq:IC}
\textrm{IC}_n=\dfrac{(x_i^\prime-x_j^\prime)}{\sqrt{t_i^\prime+t_j^\prime}}
\end{equation}
where $x_n^\prime = x_n$ and $t_n^\prime = t_n$ when $x_n$ is known with certainty (for example when node $n$ is a tip on the phylogeny associated with an empirically observed trait value) and otherwise
\begin{align}
x_n^\prime &= \dfrac{x_i^\prime/t_i^\prime+x_j^\prime/t_j^\prime}{1/t_i^\prime+1/t_j^\prime} \label{eq:ICxn}\\
 t_n^\prime &= t_n + \dfrac{t_i^\prime t_j^\prime}{t_i^\prime + t_j^\prime} \label{eq:ICtn}
\end{align}

\subsection{Maximum likelihood ancestral state reconstruction}

Consider a node $n$ connected to its parent $p_n$ by a branch of length $t_n$. The change in the value of the evolving trait along this branch, $x_n-x_{p_n}$ is normally distribution with mean zero and variance proportional to $t_n$. The likelihood of the tree with respect to some candidate ancestral state reconstruction $\textbf{X}$ is given by the product of the normal distributions associated with each branch:
\begin{equation}
\label{eq:treeML}
\textrm{L}(\textbf{X}; \mathcal{T}) \propto \prod_{n} \phi(x_n-x_{p_n};0,\sqrt{t_n})
\end{equation}
where  $\phi(x;\mu,\sigma)$ is the probability density of the normal distribution mean $\mu$ and variance $\sigma^2$. The task of maximum likelihood ancestral state reconstruction is to identify the $\hat{\textbf{X}}$ which maximizes the likelihood function in \eqref{eq:treeML}. 

Maddison (1994) notes that, in addition to the global maximum likelihood ancestral state reconstruction $\hat{\textbf{X}}$, there exists a local maximum likelihood ancestral state reconstruction $\tilde{\textbf{X}}$ such that each $\tilde{x}_n$ is the ancestral state maximizing the likelihood of the subtree rooted at node $n$. Since \eqref{eq:treeML} is a product of normal distributions, the local likelihood function at node $n$ is also a normal distribution which we denote $\textrm{N}(\tilde{\mu}_n,\tilde{\sigma}_n)$. Evidently the local maximum likelihood estimate of $\tilde{x}_n$ must be equal to $\tilde{\mu}_n$. The parameters $\tilde{\mu}_n$ and $\tilde{\sigma}_n$ for each $n$ can be estimated in a traversal from the tips of the phylogeny to its root. For each tip $t$ of the phylogeny with known trait value $x_t$, we have $\mu_t=x_t$ and $\sigma_t=0$ if the tip trait value is known with certainty 
or a nonzero value if the tip trait's distribution is known. For each internal node $n$ with two children $i$ and $j$ connected by branches of length $t_i$ and $t_j$, the local likelihood function with respect to $x_n$ is given by the product of the local likelihood functions at $i$ and $j$ suitably weighted by the branch lengths under the Brownian assumption of additive variance:
\begin{align}
\phi(x_n; \tilde{\mu}_n, \tilde{\sigma}_n) &= \phi(\tilde{\mu}_i - x_n; 0, \sqrt{\sigma_i^2 + t_i}) \;  \phi(\tilde{\mu}_j - x_n; 0, \sqrt{\sigma_j^2 + t_j}) \label{eq:localPhi1}\\
                     &= \phi\left(x_n; \dfrac{\tilde{\mu}_i (\tilde{\sigma}_j^2 + t_j) + \tilde{\mu}_j (\tilde{\sigma}_i^2 + t_i)}
 { (\tilde{\sigma}_i^2 + t_i) +  (\tilde{\sigma}_j^2 + t_j)},
 \sqrt{\dfrac{(\tilde{\sigma}_i^2 + t_i)(\tilde{\sigma}_j^2 + t_j)}{(\tilde{\sigma}_i^2 + t_i)+(\tilde{\sigma}_j^2 + t_j)}}\right) \label{eq:localPhi2}
\end{align}
from which the values of $\tilde{x}_n = \tilde{\mu}_n$ and $\tilde{\sigma}_n$ can be read off.

Similarly we denote the maximum likelihood function at node $n$ with respect to $x_n$ as $\textrm{N}(\hat{\mu}_n,\hat{\sigma}_n)$. As mentioned above, $\hat{\mu}_{root}$ = $\tilde{\mu}_{root}$ and $\hat{\sigma}_{root}$ = $\tilde{\sigma}_{root}$. For other nodes of the phylogeny for which the maximum likelihood function must be estimated, we follow a similar logic to obtain:
\begin{align}
\phi(x_n; \hat{\mu}_n, \hat{\sigma}_n) &= \phi\left(\tilde{\mu}_n - x_n; 0, \tilde{\sigma}_n\right) \phi\left(x_n- \hat{\mu}_{p_n};0,\sqrt{t_n}\right)\\
                                       &= \phi\left(x_n; \dfrac{\tilde{\sigma}_n^2 \hat{\mu}_{p_n} + t_n \tilde{\mu}_n}{\tilde{\sigma}_n^2+t_n},  \sqrt{ \dfrac{t_n \tilde{\sigma}_n^2}{t_n + \tilde{\sigma}_n^2} }  \right)
\end{align}
from which the values of $\hat{x}_n = \hat{\mu}_n$ and $\hat{\sigma}_n$ can be read off.

We note that Maddison (1994) has described a similar two-pass algorithm based on the quadratic function describing the sum of squared deviations in trait value on the phylogeny, a method yielding identical local and global ancestral state estimates resulting from somewhat more complex formulae. The distributional approach described above has the benefit of directly yielding confidence intervals based on $\hat{\sigma}_n$ for each node $n$.

\section{Results}

\subsection{Independent contrasts can be expressed in terms of the local maximum likelihood ancestral state reconstruction}

Consider Equation \eqref{eq:ICtn}. When calculating independent contrasts by traversing from the tips to the root of a phylogeny, each branch is extended by a factor of
\begin{equation}
 \dfrac{t_i^\prime t_j^\prime}{t_i^\prime + t_j^\prime}
\end{equation}
a term which is strikingly similar in form to the variance of the local maximum likelihood function at node $n$ defined in Equation \eqref{eq:localPhi2} as $\tilde{\sigma}_n^2$:
\begin{equation}
 \dfrac{(\tilde{\sigma}_i^2 + t_i)(\tilde{\sigma}_j^2 + t_j)}{(\tilde{\sigma}_i^2 + t_i)+(\tilde{\sigma}_j^2 + t_j)}
\end{equation}
Indeed if we grant that each tip $t$ of the phylogeny has a fixed point estimate of $x_t$, with $\tilde{\sigma}_t = 0$, then for any node whose children are tips we have
\begin{equation}
 \dfrac{t_i^\prime t_j^\prime}{t_i^\prime + t_j^\prime} = \dfrac{(\tilde{\sigma}_i^2 + t_i)(\tilde{\sigma}_j^2 + t_j)}{(\tilde{\sigma}_i^2 + t_i)+(\tilde{\sigma}_j^2 + t_j)}
\end{equation}
and this identity will hold for all branches since the left side of the equation, like the right side, is additive down the the phylogeny and stored as a constant factor in the transformed branch lengths (under independent contrasts) or in the variance of the local maximum likelihood function (under maximum likelihood reconstruction), resulting in the equality:
\begin{equation}
\label{tnprime_in_terms_of_tildesigma}
 t_n^\prime = t_n + \tilde{\sigma}_n^2 
\end{equation}

Note that the formula for $x_n^\prime$ given in Equation \eqref{eq:ICxn} simplifies to:
\begin{equation}
\label{eq:ICxn_simplified}
 x_n^\prime = \dfrac{x_i^\prime t_j^\prime + x_j^\prime t_i^\prime}{t_i^\prime + t_j^\prime}
\end{equation}
which is strikingly similar in form to the mean of the local maximum likelihood function at node $n$ defined in Equation \eqref{eq:localPhi2}:
\begin{equation}
\label{eq:tildemun}
 \tilde{\mu}_n = \dfrac{\tilde{\mu}_i (\tilde{\sigma}_j^2 + t_j) + \tilde{\mu}_j (\tilde{\sigma}_i^2 + t_i)}
 { (\tilde{\sigma}_i^2 + t_i) +  (\tilde{\sigma}_j^2 + t_j)}
\end{equation}
By substituting according to equation \eqref{tnprime_in_terms_of_tildesigma} we obtain:
\begin{equation}
 \tilde{\mu}_n = \dfrac{\tilde{\mu}_i t_j^\prime + \tilde{\mu}_j t_i^\prime}{t_i^\prime+t_j^\prime}
\end{equation}
Again, since $\tilde{\mu}_t = x_t^\prime$ for any tip with fixed trait value, by induction on Equation \eqref{eq:ICxn_simplified} it follows that
\begin{equation}
\label{xnprime_in_terms_of_tildemu}
 x_n^\prime = \tilde{\mu}_n
\end{equation}
for general $n$.

Equations \eqref{tnprime_in_terms_of_tildesigma} and \eqref{xnprime_in_terms_of_tildemu} provide a fundamental connection between maximum likelihood ancestral state reconstruction and independent contrasts, permitting us to represent an independent contrast at node $n$ in terms of the local maximum likelihood function. By substituting into \eqref{eq:IC} we obtain:

\begin{equation}
  \textrm{IC}_n = \dfrac{(\tilde{\mu}_i-\tilde{\mu}_j)}{\sqrt{t_i + \tilde{\sigma}_i^2 +t_j + \tilde{\sigma}_j^2 }}\label{eq:ICn_in_terms_of_tilde_mu_ij}
\end{equation}

\subsection{The sum of squared independent contrasts over a phylogeny is identical to the sum of squared deviations over a phylogeny imputed by the global maximum likelihood ancestral state reconstruction}

Since the local likelihood function at node $n$ is a normal distribution with mean $\tilde{\mu}_n$, the local maximum likelihood ancestral state estimate is $\tilde{x}_n = \tilde{\mu}_n$, a quantity which generates the following sum of squared deviations in the evolving trait on branches leading from $n$ to its children:
\begin{equation}
\label{eq:BrownianSS}
 \tilde{SS}_n = \dfrac{(\tilde{\mu}_i-\tilde{\mu}_n)^2}{\tilde{\sigma}_i^2+t_i} + \dfrac{(\tilde{\mu}_j-\tilde{\mu}_n)^2}{\tilde{\sigma}_j^2+t_j}
\end{equation}
By substituting \eqref{eq:tildemun} we obtain:
\begin{align}
 \tilde{SS}_n &= \dfrac{(\tilde{\mu}_i-\tilde{\mu}_j)^2}{t_i + \tilde{\sigma}_i^2 +t_j + \tilde{\sigma}_j^2}\\
             &= \textrm{IC}_n^2\label{eq:BrownianSS_to_IC}
\end{align}

Since, at the root, the local likelihood and sum of squares is equal to the global likelihood and sum of squares, and given that the sum of squared deviations derived from the pair of branches descending from any node is identical to the squared independent contrast, we obtain:

\begin{equation}
\sum_n \dfrac{(\tilde{\mu}_n-\tilde{\mu}_{p_n})^2}{\tilde{\sigma}_n^2+t_n} = \sum_n \dfrac{(\hat{\mu}_n-\hat{\mu}_{p_n})^2}{t_n} = \sum_n IC_n^2
\end{equation}

\subsection{Regression estimators derived from maximum likelihood ancestral state reconstruction and from independent contrasts yield identical estimates of slope and correlation coefficient for bivariate traits evolving under Brownian motion}

Reduced major axis and ordinary least squares estimators for slope and correlation coefficient depend solely on the variance and covariance of the variables subject to regression analysis. The variance of trait $\textbf{X}$ given a maximum likelihood ancestral state reconstruction $\hat{\textbf{X}}$ is given by

\begin{equation}
 \sigma_{\hat{\textbf{X}}}^2 =  \dfrac{1}{2t-2} \sum_n \dfrac{(\hat{\mu}_{\textbf{X}_n}-\hat{\mu}_{\textbf{X}_{p_n}})^2}{t_n}
\end{equation}
where the phylogeny has $t$ tips and $2t-2$ is the number of edges.
The variance based on independent contrasts is:
\begin{equation}
 \sigma_{\textrm{IC}_\textbf{X}}= \dfrac{1}{t-1}  \sum_n {{\textrm{IC}_\textbf{X}}_n}^2
\end{equation}
because two branches are consumed by each independent contrast. For this reason $\sigma_{\textrm{IC}_\textbf{X}}^2$ is exactly twice as large as $\sigma_{\hat{\textbf{X}}}$. It can be shown using Equation \eqref{eq:ICn_in_terms_of_tilde_mu_ij} and the same line of reasoning that the covariance $\sigma_{\textrm{IC}_\textbf{XY}}^2$ is also exactly twice as large as $\sigma_{\hat{\textbf{X}}\hat{\textbf{Y}}}^2$.

Due to cancelling out of the denominators in the variance terms, the correlation coefficient based on independent contrasts, $\rho_{IC}$, is identical to that based on maximum likelihood ancestral state reconstruction, $\rho_{ML}$:
\begin{equation}
 \rho_{\textrm{IC}} =\dfrac{\sigma_{\textrm{IC}_\textbf{XY}}^2}{\sigma_{\textrm{IC}_\textbf{X}} \sigma_{\textrm{IC}_\textbf{Y}}} =\dfrac{2\sigma_{\hat{\textbf{X}}\hat{\textbf{Y}}}^2}{\sqrt{2}\sigma_{\hat{\textbf{X}}} \sqrt{2}\sigma_{\hat{\textbf{Y}}}} =\rho_{ML}
\end{equation}
Similarly, the ordinary least squares regression slope estimators based on independent contrasts and maximum likelihood ancestral state reconstruction are also identical:
\begin{equation}
 \beta_{\textrm{IC}} = \dfrac{\sigma_{\textrm{IC}_\textbf{XY}}}{\sigma_{\textrm{IC}_\textbf{X}}} = \dfrac{\sqrt{2}\sigma_{\hat{\textbf{X}}\hat{\textbf{Y}}}}{\sqrt{2}\sigma_{\hat{\textbf{X}}}} =  \beta_{\textrm{ML}}
\end{equation}
as are the reduced major axis regression slope estimators:
\begin{equation}
  \beta_{\textrm{IC}} = \textrm{sign}(\rho_{\textrm{IC}}) \, \dfrac{\sigma_{\textrm{IC}_\textbf{Y}}}{\sigma_{\textrm{IC}_\textbf{X}}} = \textrm{sign}(\rho_{\textrm{ML}}) \, \dfrac{\sqrt{2}\sigma_{\hat{\textbf{Y}}}}{\sqrt{2}\sigma_{\hat{\textbf{X}}}} =\beta_{\textrm{ML}}
\end{equation}

\subsection{Independent contrasts for Brownian motion with a directional tendency}

The standard Brownian motion model of continuous character evolution has zero mean such that the expected value of a trait after a period of evolution of duration $t$ is equal to the value of the trait prior to the period of evolution. A directional tendency to the evolutionary process can be modelled in terms of a nonzero mean $M$, such that the expected value of a trait after a period of evolution of duration $t$ is equal to $tM$. By modifying Equation \eqref{eq:localPhi2} appropriately it is trivial to incorporate the directional tendency into the model described previously. In a traversal from the tips of the tree to the root we define the local likelihood function for each internal node:
\begin{equation}
\label{tildemun_directional}
 \tilde{\mu}_n = \dfrac{(\tilde{\mu}_i-t_iM ) (\tilde{\sigma}_j^2 + t_j) + (\tilde{\mu}_j -t_jM )(\tilde{\sigma}_i^2 + t_i)}
 { (\tilde{\sigma}_i^2 + t_i) +  (\tilde{\sigma}_j^2 + t_j)}
\end{equation}
\begin{equation}
 \tilde{\sigma}_n=\sqrt{\dfrac{(\tilde{\sigma}_i^2 + t_i)(\tilde{\sigma}_j^2 + t_j)}{(\tilde{\sigma}_i^2 + t_i)+(\tilde{\sigma}_j^2 + t_j)}}
\end{equation}
and calculate a phylogenetically independent contrast incorporating directional tendency:
\begin{equation}
 IC_n^2 = \dfrac{(\tilde{\mu}_i-\tilde{\mu}_n-t_i M)^2}{\tilde{\sigma}_i^2+t_i} + \dfrac{(\tilde{\mu}_j-\tilde{\mu}_n-t_jM)^2}{\tilde{\sigma}_j^2+t_j}\label{eq:directionalPC}
\end{equation}
Substituting \eqref{tildemun_directional} into \eqref{eq:directionalPC} we obtain:
\begin{equation}
 IC_n = \dfrac{(\tilde{\mu}_i-\tilde{\mu}_j)- M (t_i - t_j)}{\sqrt{t_i+\tilde{\sigma}_i^2+t_j+\tilde{\sigma}_j^2}}\label{eq:directionalIC}
\end{equation}

For the sake of completeness, we here also define the maximum likelihood function for each node, which can be calculated in a second traversal from the root of the phylogeny to its tips in order to obtain maximum likelihood ancestral states under a directional tendency:
\begin{equation}
 \hat{\mu}_n = \dfrac{\tilde{\sigma}_n^2 (\hat{\mu}_{p_n} + t_n M) + t_n \tilde{\mu}_n}{\tilde{\sigma}_n^2+t_n} 
\end{equation}
\begin{equation}
 \hat{\sigma}_n = \sqrt{ \dfrac{t_n \tilde{\sigma}_n^2}{t_n + \tilde{\sigma}_n^2} }
\end{equation}

Given some value of $M$ it is thus possible to calculate a set of phylogenetically independent contrasts for a trait evolving with directional tendency. The value of $M$ is typically not known, but is easily estimated from the data by conducting a linear search to identify the $\hat{M}$ which minimizes the sum of squared contrasts or maximizes the global likelihood of the model.

\end{document}